%% file: main.tex
\documentclass[fleqn,usenatbib]{mnras}

\usepackage{newtxtext,newtxmath}

\usepackage[T1]{fontenc}
\usepackage{ae,aecompl}

\usepackage{graphicx}	
\usepackage{amsmath}	
\usepackage{amssymb}	
\usepackage{url}
\usepackage{xspace}
\usepackage{upgreek}
\usepackage{nicefrac}
\usepackage{bm}
\usepackage{ulem}
\usepackage{xspace}

\input{mycommands.tex}

\input{myacro_arxiv.tex}

\graphicspath{{./}{figures/}}

\title[UHE gamma rays from the Crab Nebula]{Detection of UHE gamma rays from the Crab Nebula: Physical Implications}

\author[D. Khangulyan et al.]{
Dmitry Khangulyan,$^{1}$\thanks{E-mail: d.khangulyan@rikkyo.ac.jp}
Masanori Arakawa,$^{1,2}$
Felix Aharonian$^{3,4,5}$
\\
$^{1}$\rikkyo\\
$^{2}$Astrophysical Big Bang Laboratory, RIKEN, Saitama 351-0198, Japan\\
$^{3}$\dias\\
$^{4}$\mpik\\
$^{5}$\mephi\\
}

\date{Accepted XXX. Received YYY; in original form ZZZ}

\pubyear{2019}

\begin{document}
\label{firstpage}
\pagerange{\pageref{firstpage}--\pageref{lastpage}}
\maketitle

\begin{abstract}
  The Crab Nebula is an extreme particle accelerator boosting the energy of electrons up to a few PeV
  ($10^{15} \ \rm eV$), close to the maximum energy allowed by theory.  The physical conditions in the acceleration site
  and the nature of the acceleration process itself remain highly uncertain. The key information about the highest
  energy accelerated particles is contained in the synchrotron and inverse Compton (IC) channels of radiation at
  energies above 1~MeV and 100~TeV, respectively. The recent report of detection of ultra-high energy gamma-ray signal
  from the Crab Nebula up to 300~TeV allows one to determine the energy distribution of the highest energy electrons
    and to derive the magnetic field strength in the acceleration region, \(B\leq120\rm\,\upmu G\), in a parameter-free
    way.  This estimate brings new constraints on the properties of non-thermal particle distributions and puts
  important constraints on the MHD models for the Crab Nebula, in particular on the feasible magnetization and
  anisotropy of the pulsar wind. The calculations of synchrotron and IC emission show that future observations with
  instruments allowing detection of the Crab Nebula above 300~TeV and above 1~MeV will clarify the conditions allowing
  acceleration of electrons beyond PeV energies in the Crab Nebula. In particular, one will (1) verify the hypothetical
  multi-component composition of the electron energy distribution, and (2) determine the magnetic field strength in the
  regions responsible for the acceleration of PeV electrons.
\end{abstract}




\section{Introduction}\label{sec:intro}
The rotation-powered pulsars initiate high energy gamma-radiation in three physically distinct regions called (i) pulsar
magnetosphere, (ii) relativistic electron-positron wind, (iii) \ac{pwn}.  The cold ultrarelativistic pulsar wind
originating from the pulsar magnetosphere and carrying almost the entire rotational energy of the pulsar, eventually
terminates resulting in the formation of the nonthermal synchrotron and \ac{ic} nebula.  The nonthermal emission of
\acp{pwn} is caused by interactions of relativistic electrons accelerated at the relatively compact regions associated
with the termination of the wind. However, because of the diffusive and advective propagation of electrons, the
nonthermal radiation typically extends to distances tens of parsecs.  The energy density of the magnetic field in most
of \acp{pwn} is comparable to the ambient radiation fields. Thus, the energy of relativistic electrons is shared between
the synchrotron and \ac{ic} channels of radiation in fair fractions making \acp{pwn} not only effective electron
accelerators but also very effective gamma-ray emitters \citep{1995NuPhS..39A.193A,1997MNRAS.291..162A}.

The Crab Nebula is unique, but not an archetypical (as often claimed in the literature) representative of \acp{pwn}. Its
pulsar is much more powerful than the pulsars of most of other \acp{pwn} and, surprisingly, the Crab Nebula is unusually
compact. The magnetic field in the Crab Nebula exceeds, by an order of magnitude or more, the typical
($\sim 10\rm\, \upmu G$) strength of the magnetic field in other \acp{pwn}. Correspondingly the energy density of the
magnetic field exceeds by two or three orders of magnitude the radiation fields, thus violating the balance between the
synchrotron and \ac{ic} radiation channels.  This makes the Crab Nebula a rather inefficient gamma-ray emitter.  The low
\ac{ic} radiation efficiency is compensated by the vast rotational power of the Crab pulsar. Therefore, despite the low
efficiency, the Crab Nebula remains a very strong gamma-ray source.

The Crab Nebula is characterized by a very broad \ac{sed} that spans over 20 decades, from MHz radio wavelengths to
\ac{uhe} gamma-rays. Another unique feature of the Crab Nebula is the extension of its synchrotron radiation to MeV,
and, during the flares, to GeV energies \citep[see][and references therein]{flare_review,2017EPJWC.13603003Z}, implying
that we deal with an extreme accelerator in which the acceleration of electrons proceeds close to the maximum possible
rate allowed in ideal \ac{mhd} configurations \citep{1996ApJ...457..253D,2002PhRvD..66b3005A,2010MNRAS.405.1809L}.

Thanks to its proximity, \(d\simeq2\rm\,kpc\), and the high spin-down luminosity of the pulsar,
\(L\sub{sd}\simeq5\times10^{38}\ergs\), the Crab Nebula can be studied in great details \citep[see][and references
therein]{Hester2008}. It contains several spectral features and demonstrates complex energy-dependent morphology
\citep[see, e.g.,][]{Hester2008,Weisskopf2000,Nustar,2019ApJ...875..123Y,2019arXiv190909494H}.  The size of the Crab
Nebula depends on photon energy, which most likely is caused by the energy-dependent cooling times of the parent
electrons. This suggests that the particle acceleration occurs in the inner part of the nebula where relativistic pulsar
wind terminates.  This conclusion contains, however, caveats.  Despite intensive theoretical studies and numerical
simulations, it is still not clear how the particle acceleration operates at the relativistic shock waves
\citep[see][and references therein]{2015SSRv..191..519S}.  It is not clear which magnetic field strength and its
configuration are required for efficient particle acceleration. Non-thermal particles seen in the Crab Nebula may
originate from different acceleration sites, not necessary associated with the pulsar wind \ac{ts}. Even if the \ac{ts}
is the major accelerator in the Crab Nebula, the physical conditions should vary considerably depending on the specific
region of the pulsar wind \ac{ts}, as the \ac{ts} is expected to have a complex non-spherical shape
\citep{2002AstL...28..373B,2002MNRAS.329L..34L}. Thus, it cannot be excluded that several distinct population of
particle, which are possibly accelerated by different mechanisms and/or under different condition, coexist in the Crab
Nebula \citep{Aharonian1998,2019MNRAS.489.2403L}.  Only precise simulations, together with detailed observations, can
help to localize the principal regions of particle acceleration \citep{2015MNRAS.449.3149O}.

It has been argued \citep{Atoyan1996} that at least two distinct population of electrons should be invoked to explain
the broad-band non-thermal emission of the Crab Nebula. The ``radio electrons'' are responsible for the MHz and GHz
synchrotron emission, while the \ac{ic} up-scattering off soft photons by these electrons results in the gamma-ray
component at GeV energies. The modelling of the radio morphology does not constrain the origin of the radio-emitting
electrons in the Crab Nebula, and (re)acceleration of these particle in the bulk of the nebula cannot be excluded
\citep{2014MNRAS.438.1518O}.

Multi TeV electrons are classified as ``wind electrons'' \citep{Atoyan1996}. Assuming that the emitting electrons are
injected at the pulsar wind \ac{ts} and get advected with the down-stream flow, the extension of the Crab Nebula seen in
UV, X-ray, and TeV energy bands \citep{Nustar,2019arXiv190909494H} can be adequately reproduced.  We may conclude that
the \ac{ts} plays a principal role in the acceleration of multi-TeV electrons and their injection into the nebula.

Formally, a power-law spectrum of ``wind electrons'' with an exponential cut-off at PeV energies and a hardening (or a
break) at sub-TeV energies, together with the additional ``radio electrons'' can explain the broad-band spectrum of the
Crab Nebula over 20 decades.  The nonthermal emission of the Crab Nebula has been studied within a simple one-zone
approach or using a more appropriate MHD treatment of particle transport in the nebula
\citep{Kennel1984b,1992ApJ...396..161D, Atoyan1996}. As long as it concerns the spectral fits, the conclusions of both
approaches are similar \citep[see, e.g.,][]{Meyer2010}. Namely, the cut-off energy in the spectrum of the ``wind
electrons'' is close to one PeV and the spectrum should continue up to \(\simeq5\)~PeV, while the average nebular
magnetic field is constrained within \(100\) and \(300\rm\,\upmu G\).

The analysis of the acceleration process responsible for the ``wind electrons'' poses a few conceptual questions, for
example, regarding the maximum attainable electron energy. The magnetic field at the accelerator makes electrons to lose
their energy due to synchrotron cooling.  If the synchrotron losses dominate as a radiative cooling channel, the product
of the maximum attainable energy and the square root of the magnetic field strength,
\(E\sub{max} B\sub{acc}^{\nicefrac12}\), can be taken as a measure of the accelerator efficiency.  The extension of the
electron spectrum to PeV energies in the magnetic field exceeding $100\rm\, \upmu G$, tells us that in the Crab Nebula
we deal with an acceleration efficiency approaching to the ideal \ac{mhd} limit, i.e., the strength of the accelerating
electric field gets close to the strength of the magnetic field.

Another important issue of the spectral modelling of the Crab Nebula is related to the so-called Crab flares -- intense
flashes of synchrotron radiation extending to GeV energies \citep{Crab_flare_1st,2011Sci...331..736T}.  The short
variability time scale and the distinct spectral shape of the flaring component require a magnetic field which is
significantly larger than the average nebular field, \(B\geq1\rm\,mG\), a non-ideal \ac{mhd} configuration, and/or
relativistic motions with large bulk Lorentz factor \citep[see][and references therein]{flare_review}. Even under these
extreme conditions, the Crab flares require the presence of electrons with energy exceeding a few PeV. The Crab flares
are likely to be formed in a region(s) quite different from the site(s) where the steady component of radiation is
produced.  But in both cases, the acceleration of electrons proceeds with efficiency close or even beyond the ideal
\ac{mhd} limit.

The tough efficiency requirement to the acceleration of the ``wind electrons'' can be significantly relaxed assuming
that another component of electrons is responsible for the multi-MeV gamma-rays. This hypothesis is supported by a
non-smooth transition between the \ac{integral} and \ac{comptel} spectra. If this spectral feature is real, then
  invoking an additional component of electrons with a hard energy distribution  one can better reproduce the spectral structure at multi-MeV
  energies \citep{Aharonian1998}. The currently available data do not exclude a significantly stronger magnetic field
in the regions where this hypothetical hard distribution is localized \citep{Aharonian1998}.  Therefore, the
\ac{ic} emission  produced by electrons from the hard component  could be suppressed compared to the \ac{ic}
radiation associated with the standard ``wind electron'' component.  The synchrotron radiation alone does not allow
  one to infer information about the electrons and magnetic fields in this most critical, for highest
energy electrons, MeV/GeV energy region. For any reasonable magnetic field, the MeV synchrotron radiation is produced by
multi-hundred TeV electrons. Therefore, the UHE gamma-rays produced by the same electrons through the \ac{ic} scattering
contains information which can shed light on both electron energy distribution and magnetic field strength.

In this regard, the recently reported detection of fluxes gamma-rays above 100 ~TeV \citep{tibet,hawc} are of great
interest. These measurements significantly extend the spectrum, which previously was measured up to \(80\rm\,TeV\)
\citep{hegra}.  The emission reported by \ac{tibet} and \ac{hawc} collaborations is produced by multi-hundred TeV
electrons scattering off the \ac{cmbr} allowing robust constraints on the parent electron spectrum and the magnetic
  field strength in the region of acceleration.  Below we discuss some implications of the new \ac{uhe} gamma-ray
measurements in the context of the highest energy electrons accelerated in the Crab Nebula.

\section{ UHE Particles in the Crab Nebula}
\subsection{Transport}
While high-energy particles in \acp{pwn}  lose their energy due to synchrotron, \ac{ic}, and adiabatic losses, in the \ac{uhe} domain synchrotron losses are expected to dominate. The synchrotron cooling  time is:
\be\label{eq:syn_cooling}
t\sub{syn}\simeq 4\times10^{5}E\sub{PeV}^{-1}B_{\rm mG}^{-2}\rm\,s\, ,
\ee
were \(E\sub{PeV}\) and \(B_{\rm mG}\) are the  electron energy and magnetic field strength in units of PeV and mG, respectively. The magnetic field in the nebula originates in the pulsar.  Its characteristic strength is determined by the distance from the pulsar to the point  where the pulsar wind terminates \citep{1974MNRAS.167....1R,Kennel1984b}.
In \acp{pwn}, the down-stream magnetic field is changed on scales comparable to the \ac{ts} radius \citep[see, e.g.,][]{Kennel1984b}. It is convenient to introduce the so-called dimensionless advection distance:
\be \label{eq:ad_distance_dimensionless}
\lambda\syn=\frac{t\syn V\sub{f}}{R\sub{ts}}=\frac{10^{-2}}{B_{\rm mG}^2E\sub{PeV}} \left(\frac{V\sub{f}}{0.3}\right)\left(\frac{R\sub{ts}}{0.1\rm\ pc}\right)\,,
\ee
where \(V\sub{f}\) is flow velocity in the units of the speed-of-light.
The above equation suggests that the emission of \ac{uhe} \(>100\rm\,TeV\) electrons provides good probe for  physical conditions in the acceleration region.
\subsection{Emission}

The synchrotron emission of PeV electrons in a magnetic field of mG scales  appears in the gamma-ray energy band:
\be
\hbar\omega\sub{syn}\simeq60 E\sub{PeV}^2B_{\rm mG}\rm\,MeV\,,
\ee

The synchrotron radiation alone does not provide independent information about the electrons and the magnetic field.  The \ac{ic} component of radiation of the same electrons allows us
to disentangle the strength of the magnetic field and the energy parent electrons.

In the Crab Nebula, several photon fields serve as targets for the \ac{ic} emission \citep{Atoyan1996}. Three dominant \ac{ic} components are contributed by the  \ac{fir}, \ac{cmbr}, and synchrotron photons. Up-scattering of synchrotron photons through the \ac{ssc} channel provides the major contribution at TeV energies. The synchrotron target is however characterized by a relatively high photon energy, therefore in the UHE band the \ac{ssc} process is significantly suppressed because of the \ac{kn} effect. The \ac{kn} effect becomes substantial when
\be\label{eq:KN}
\ve\geq\ve\sub{kn}=3\times10^{-4}E\sub{PeV}^{-1}\rm\,eV\,.
\ee
The flux from the Crab Nebula at these energies is about \(10^{-10}\flux\). For the radius of \(R\sub{n}\simeq1\rm\,pc\), the energy density of the synchrotron photons  which are up-scattered in the Thomson regime, is \(\sim10^{-2}\rm\,eV\,cm^{-3}\), significantly below the energy density of the Galactic background photon fields.

\begin{figure}
  \includegraphics[width=\columnwidth]{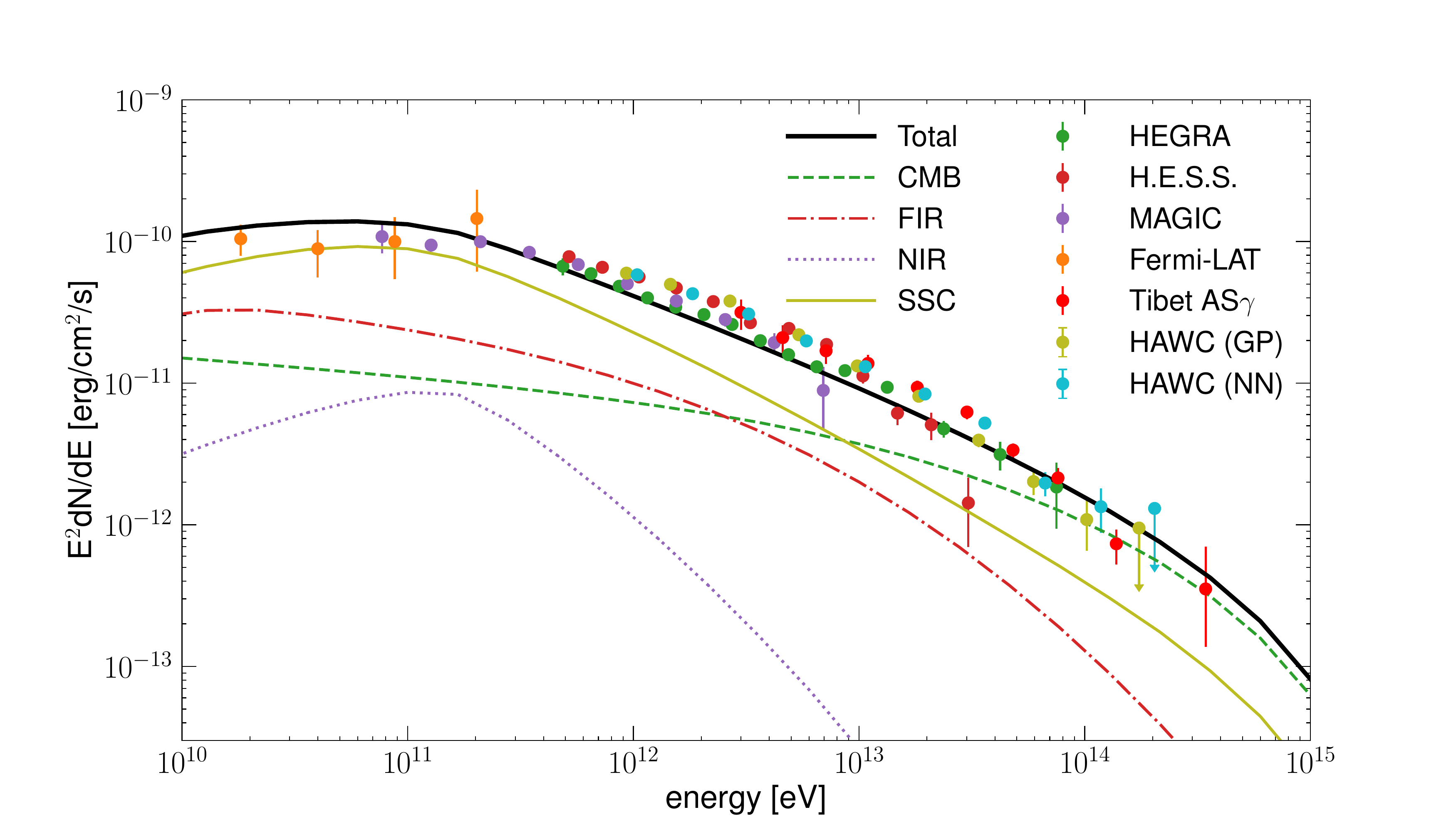}
  \caption{Gamma-ray spectrum of the Crab Nebula.
  The gamma-ray data are taken from \ac{fermi} \citep{Rolf2012}, \ac{hegra} \citep{hegra}, H.E.S.S. \citep{hess}, MAGIC \citep{magic}, \ac{tibet} \citep{tibet}, \ac{hawc} \citep{hawc}. \ac{ic} spectra produced on four different photon fields: \ac{ssc} (solid line), \ac{cmbr} (dashed line), \ac{fir} (dash-dotted line), and \ac{nir} (dotted line) are shown together with their summation (thick solid line).
  \label{fig:IC_components}}
\end{figure}

If the target is a diluted Plankian radiation, the limit given by Eq.~(\ref{eq:KN}) for \(100\rm\,TeV\) electrons corresponds to the temperature approximately of
\(10\rm \,K\). This implies that all background fields, except the \ac{cmbr}, are up-scattered deep in the
\ac{kn} regime. Thus,  the \ac{uhe} emission reported by the  \ac{tibet} and \ac{hawc} collaborations is
predominately contributed by the \ac{cmbr} photons. This conclusion is supported by  numerical calculations shown in Fig.~\ref{fig:IC_components}.
The calculations have been  performed in framework of one-zone approximation using the package {\tt{naima}} \citep{naima}.
The analytic approximation propsed in \citet{2010PhRvD..82d3002A} was used for computing the synchrotron emission.
For the SSC  component of radiation we used the IC cross-section averaged over the scattering angles \citep{1981Ap&SS..79..321A},  and the \ac{ic} spectra generated on the Galactic background fields were calculated with an analytic approximation by \citet{2014ApJ...783..100K}.
 The electron distribution was assumed to be a broken power-law with an exponential cutoff:
 \be\label{eq:electrons}
 \begin{split}
\dif[E]{N}=A\exp&{\left[ -\left(\frac{E}{E_{\rm cut}}\right)^2 \right]} \times\\
&\times\left\{
  {
    \begin{matrix}
      \left(\frac{E}{1~\rm TeV}\right)^{-\alpha_1}&{\rm if}&E<E_{\rm br}\,,\\
      \left(\frac{E_{\rm br}}{1~\rm TeV}\right)^{\alpha_2-\alpha_1}\left(\frac{E}{1~\rm TeV}\right)^{-\alpha_2} &{\rm if}&E>E_{\rm br}\,.
    \end{matrix}
  }
\right.
\end{split}
\ee
Here $E$ is electron energy, $A$ is the normalization constant for the  electron spectrum, $E_{\rm cut}$ is the cutoff energy, and $E_{\rm br}$ is power-law break energy.
We adopted the following parameters:
$E_{\rm cut}=1.863~\rm PeV$, $E_{\rm br}=0.265~\rm TeV$, $\alpha_1 = 1.5$, $\alpha_2 = 3.233$ and $B=125~\rm{\upmu G}$.

The \ac{ic} spectra generated by four different photon fields,  \ac{cmbr}, \ac{fir} (a graybody distribution with temperature $T\sub{fir}=$ 70 K and energy density $U\sub{fir}=$ 0.5 $\rm{eV~cm^{-3}}$), \ac{nir} ($kT\sub{nir}=$ 5000 K, $U\sub{nir}=$ 1 $\rm{eV~cm^{-3}}$), and
synchrotron photons, are shown  in Fig.~\ref{fig:IC_components}.
We assumed that the synchrotron target is homogeneously generated in the nebula, and the volume averaged density of the \ac{ssc} target is enhanced by a factor of \(2.24\) as compared to the boundary region of the nebula \citep[see][for details]{Atoyan1996}.

\begin{figure}
  \includegraphics[width=\columnwidth]{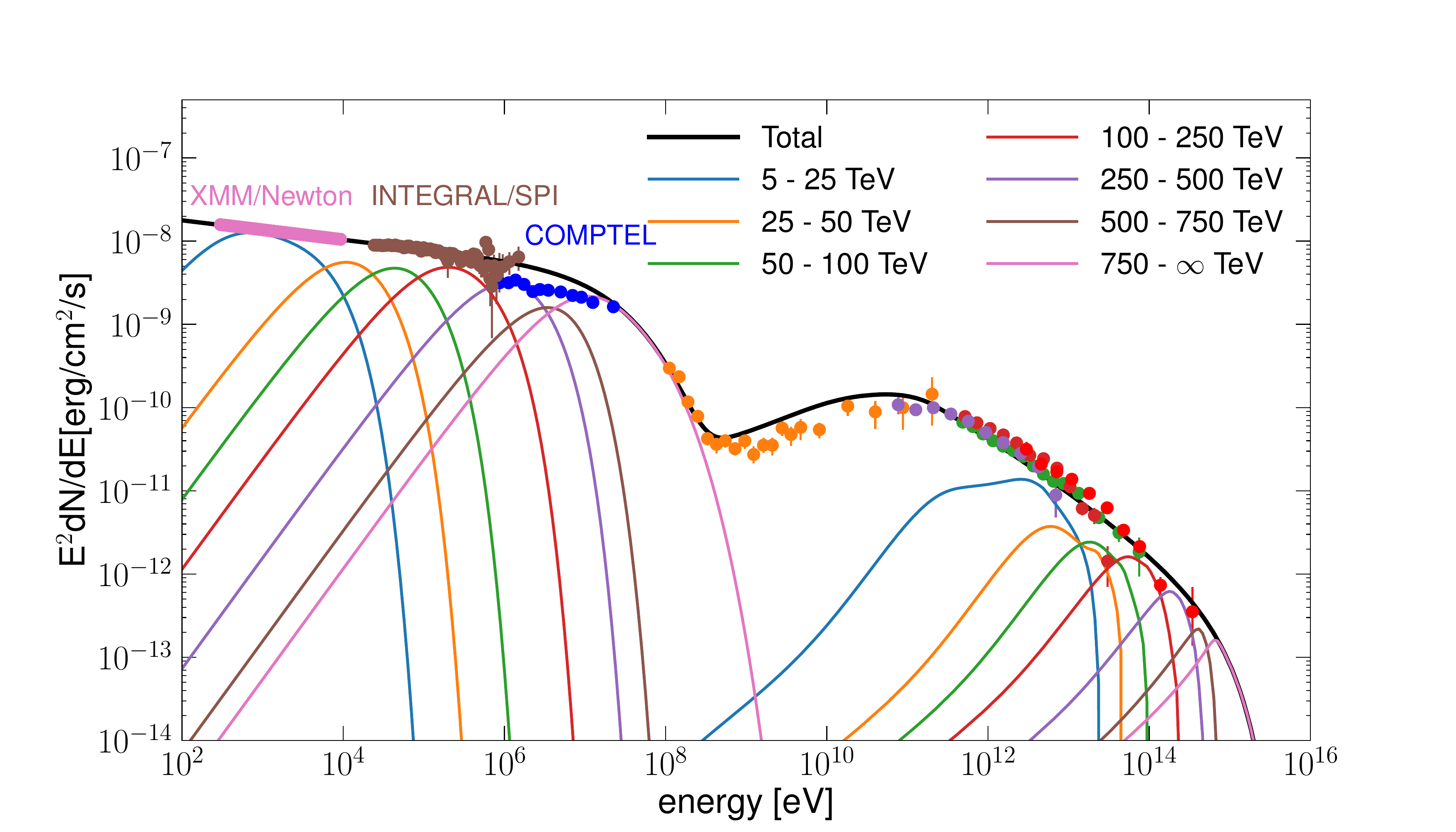}
  \caption{Non-thermal emission computed with a one-zone model (thick solid line). Synchrotron and \ac{ic} emission by electrons from several energy ranges, \(5\)~--~\(25\)~TeV, \(25\)~--~\(50\)~TeV, \(50\)~--~\(100\)~TeV, \(100\)~--~\(250\)~TeV, \(250\)~--~\(500\)~TeV, \(500\)~--~\(750\)~TeV, \(\geq750\)~TeV are shown with thin solid lines. The magnetic field was assumed to have a strength of \(B=125\rm\,\upmu G \).
  In addition to the gamma-ray data shown in Fig.~\ref{fig:IC_components}, the following X-ray and soft gamma-ray measurements are shown {\it{XMM-Newton}} \citep{xmm}, {{\it INTEGRAL}}/SPI \citep{spi}, and {\it{COMPTEL}} \citep{Crab_comptel}.
  \label{fig:IC_by_different_electrons}}
\end{figure}

Remarkably, since \ac{uhe} gamma rays  are predominantly produced at scatterings on  \ac{cmbr} photons with precisely known temperature, the spectrum  and the total energy of parent electrons can be robustly derived.  Using the analytical presentations from \citet{2014ApJ...783..100K}, one can conclude the highest energy part of the spectrum reported by \ac{tibet}, \(\hbar\omega\simeq300\rm\,TeV\) requires electrons of energy up to \(0.8\rm\,PeV\).  The corresponding numerical calculations are presented in Fig.~\ref{fig:IC_by_different_electrons}, where the synchrotron and \ac{ic} emission by electrons with energy limited to several energy intervals are shown. The overall energy distribution of electrons is assumed to obey Eq.~\eqref{eq:electrons}.

\begin{figure}
\includegraphics[width=\columnwidth]{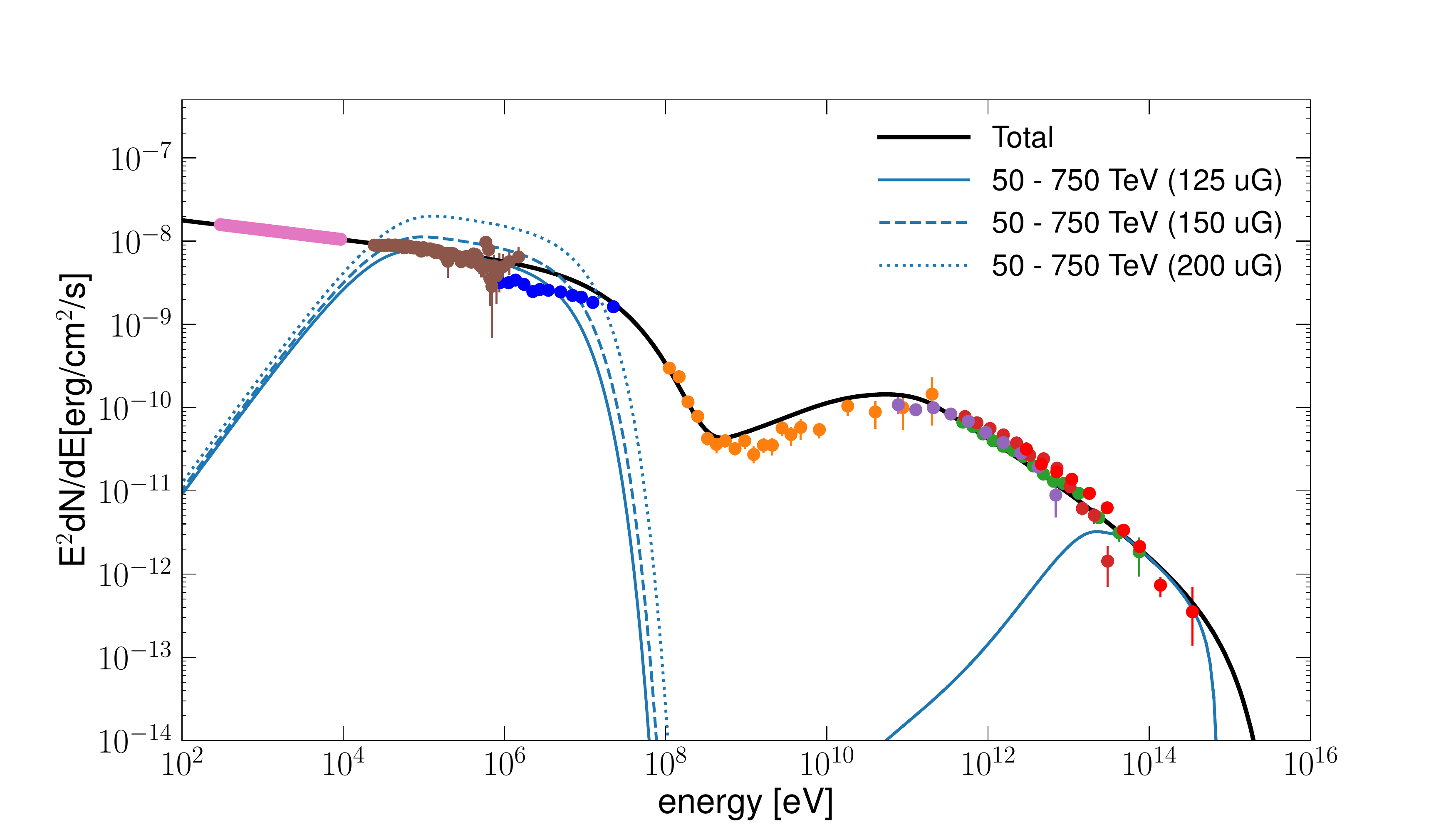}
  \caption{Computed synchrotron emission from electrons with energy in the range from \(50\) to \(750\rm\,TeV\) for three different magnetic field strengths, \(B=125\) (thin solid line), \(150\) (dashed line), and \(200\rm\,\upmu G\) (dotted line). The electron energy distribution was kept unchanged to satisfy the \ac{tibet} measurements. The synthetic non-thermal spectrum is shown with thick solid line. For the origin of the shown data points see Figs.~\ref{fig:IC_components} and \ref{fig:IC_by_different_electrons}.  \label{fig:syn_different_bfield}}
\end{figure}

It is seen that the \ac{tibet} measurements constrain the electrons in the range \(50-750\rm\,TeV\), in a parameter-free
way.  The calculation of the synchrotron emission from these particles requires
additional assumptions on the strength and possible distribution of the magnetic field. In
Fig.~\ref{fig:syn_different_bfield}, we show the synchrotron emission from \(50-750\rm\,TeV\) electrons for three
different strengths of the magnetic field as obtained in the framework of a one-zone model (all the remaining model assumptions are the same as in Fig.~\ref{fig:IC_by_different_electrons}). One can see that the
synchrotron emission of \mbox{\(\sim300\rm\,TeV\)} electrons (derived model-independently from the \ac{tibet}
measurements), violates the flux level in the MeV band measured with \ac{integral} and \ac{comptel}, unless
\(B\lesssim125\rm\,\upmu G\).  To illustrate that these data indeed constraint the strength of the magnetic field with very high accuracy, in Appendix \ref{sec:MCMC} we present a \ac{mcmc} simulation of the \ac{integral} and gamma-ray (with energy above \(10\rm\,TeV\) that include the \ac{hegra} , H.E.S.S.\xspace and \ac{tibet} measurements) spectra with {\tt naima} \citep{naima}. Adopting a power-law with exponential cutoff energy distribution of the emitting particles (which is a rather good approximation for the relatively narrow relevant  energy range), the \ac{mcmc} simulations require the magnetic field strength to be in the range \(B=118_{-2}^{+3}\rm\,\upmu G\), which is consistent with a less sophisticated estimate  shown in Fig.~\ref{fig:syn_different_bfield}.

\section{Discussion}
\subsection{Acceleration of UHE electrons}

There is an important question related to the radiation models for the Crab Nebula: do these studies allow defining the
strength of the magnetic field at the acceleration site?  As synchrotron emission components depend on a quantity
\(E B^{\nicefrac12}\), there is an ambiguity between the particle energy and the magnetic field strength. Thus, the
synchrotron spectrum alone does not define the magnetic field strength. Detection of the variability time-scale in the case of the GeV flares allows obtaining a relatively robust estimate for the magnetic field strength. The emission produced by the ``wind electrons'' is steady, however one can register the \ac{ic} component emitter by the same particles.  It means that that is \ac{ic} emission that in fact constraints the strength of
the magnetic field in the Crab Nebula. Non-thermal particle can escape from the acceleration
site and produce their emission in other parts of the source. The physical conditions at the emission  site
may differ strongly from those of the acceleration site.
If for some particle the cooling time is significantly longer then the acceleration time (i.e., the time for which the particle was confined in the accelerator), then its emission allows
probing the averaged physical conditions in the source, but not in the accelerator.
As high-energy particles are characterized by shorter cooling time, one can constrain the accelerator magnetic field by measuring the synchrotron and \ac{ic} of the highest energy particles in the source.

As argued above, the study of the synchrotron and \ac{ic} radiation components associated with \ac{uhe} electrons provide a powerful tool to constrain the magnetic field at the acceleration site. Indeed,  these electrons are expected (i) to lose their energy quickly, thus do not propagate far away from the acceleration site and (ii) interact predominately with \ac{cmbr} with precisely known distribution and energy density. The recent detection of UHE gamma-rays from the Crab Nebula beyond 100 TeV \citep{tibet} resolves the
 ``electron energy~--~magnetic field strength'' ambiguity characterizing the synchrotron channel. This detection allows us to robustly derive the strength of the magnetic field in the region responsible for the acceleration of multi-hundred TeV electrons,  \(B\leq120\rm\,\upmu G\), in a parameter-free way (see Appendix \ref{sec:MCMC}).

The obtained limit does not concern the acceleration of PeV electrons. The \ac{ic} emission of PeV electrons should appear in the spectrum above \(400\rm\,TeV\). Currently, no measurements are available in this energy band. However, it is expected that \ac{lhaaso}, a new powerful cosmic ray facility,  will soon probe the gamma-ray spectrum of the Crab Nebula in this energy interval \citep[see][and references therein]{2019ICRC...36..693H}. Remarkably, the attenuation of such energetic photons from the Crab due to the interactions with diffuse galactic radiation fields is expected to be not significant  \citep{2016PhRvD..94f3009V}.

\begin{figure}
  \includegraphics[width=\columnwidth]{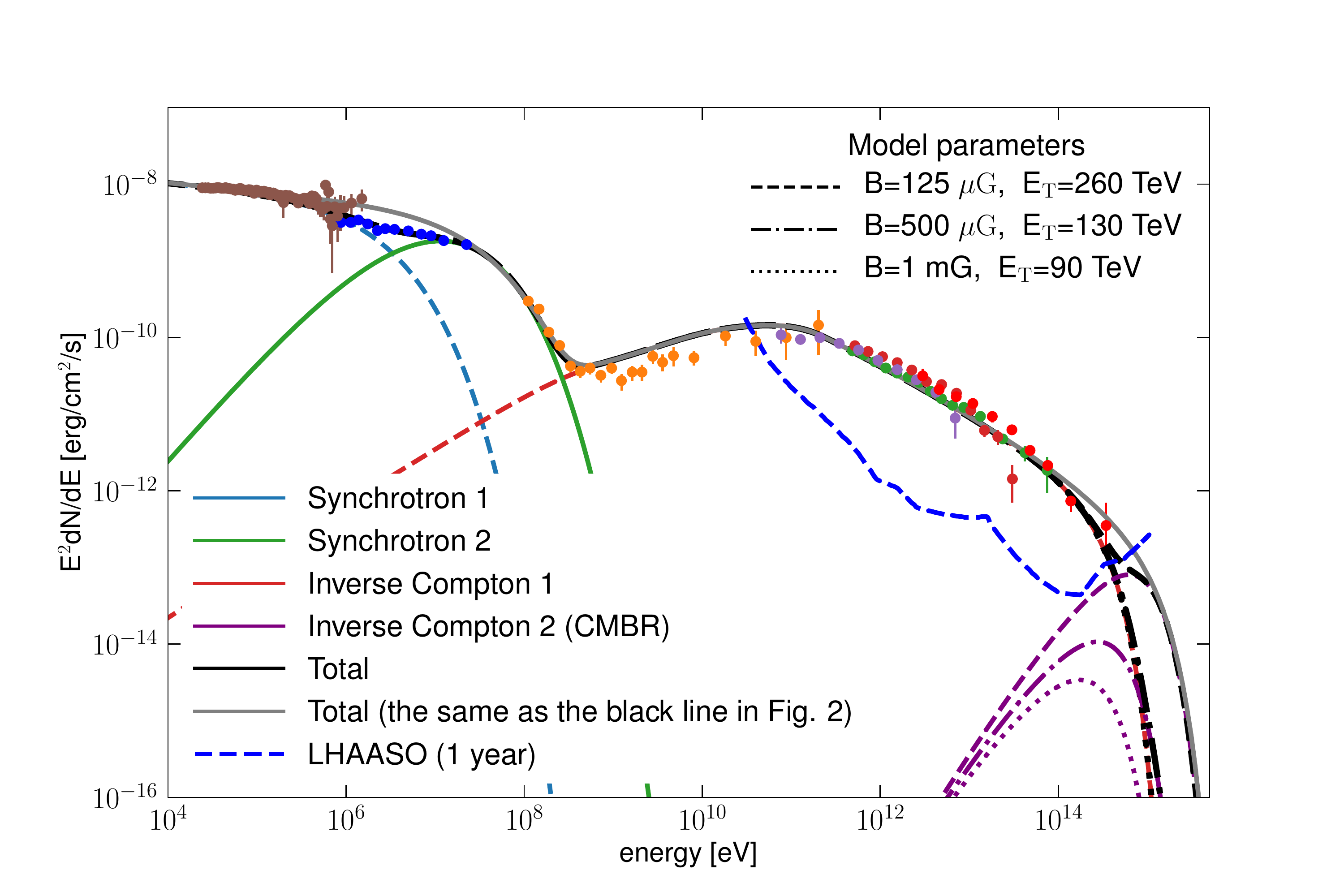}
  \caption{Computed synchrotron and \ac{ic} emission components produced by two populations of electrons: a power-law with exponential cut-off energy distribution in \(B=125\rm\,\upmu G\) magnetic field, and a relativistic Maxwellian in \(B=125\), \(500\), and \(1000\rm\,\upmu G\) fields.   For the origin of the shown data points see Figs.~\ref{fig:IC_components} and \ref{fig:IC_by_different_electrons}. \label{fig:maxwell}}
\end{figure}

The synchrotron spectrum of the Crab Nebula extends to energies beyond 100 MeV and requires electrons with energy up to
several PeV.  This emission component is conventionally associated with the ``wind electrons'', however the gamma-ray
spectrum in the 1 to 100 MeV band is not smooth. It shows a structure which can be better reproduced by two different
populations of UHE electrons described by power-law distribution with an exponential cut-off
(\(E_{\rm cut}=500\rm\,TeV\)) and by  a hard energy distribution peaking at higher energies
\citep{Aharonian1998}. The two considered electron populations could be accelerated in different regions through
  different acceleration processes. The superposition of emission of these two components is demonstrated in
  Fig.~\ref{fig:maxwell}.  Similarly to \citet{Aharonian1998} we approximated this additional component with a
  ultrarelativistic Maxwellian distribution, although we note that this is just a formal approximation, and it does not
  imply any underlying assumptions regarding the nature of this component \citep[see, however,][]{1987Ap.....26..318A}.
  For example, \citet{2014ApJ...783L..21S} have shown that a hard distribution of non-thermal particle can be formed by
  magnetic reconnection in highly magnetized environments. \citet{2019MNRAS.489.2403L} suggested that an electron
  component accelerated by magnetic field reconnection operating in the bulk of the nebula \citep[see,
  e.g.,][]{2013MNRAS.428.2459K} might be responsible for the dominant radio and soft gamma-ray emission detected from
  the Crab Nebula. Magnetic reconnection is considered as a feasible mechanism to power the Crab flares
  \citep{2012ApJ...754L..33C,2013ApJ...770..147C,2018JPlPh..84b6301L}, thus the particles producing the steady MeV and
  flaring GeV synchrotron emission may have a common origin \citep[see the discussion in][]{2019MNRAS.489.2403L}. As it
  is shown below the counterpart \ac{ic} emission may provide important information to test this possibility.

  We show in Fig.~\ref{fig:maxwell} synthetic spectra computed for three different strengths
  of the magnetic field in region where the hard high energy distribution is localized: \(B_2=125\), \(500\), and
  \(1000\rm\,\upmu G\).  As it can be seen by a suitable choice of the temperature parameter (\(E\sub{t}=260\), \(130\),
  and \(90\rm\,TeV\), respectively), one can get identical synchrotron spectra. In contrast, the \ac{ic} spectra show
  important differences (see Fig.~\ref{fig:maxwell}).

First of all, if the MeV spectral feature is real, one should expect a significantly smaller flux at
\(300\rm\,TeV\). The expected difference is comparable with the \ac{tibet} error, so presently we cannot make any
quantitative statement. However, the future measurements \ac{lhaaso} should allow distinguishing between these two cases (shown with black and gray lines in Fig.~\ref{fig:maxwell}). To illustrate that, we show in Fig.~\ref{fig:maxwell} the \ac{lhaaso} sensitivity expected for one-year exposure
\citep{2019arXiv190502773B}. We also note that important new information can be obtained, as well, in the MeV energy band,
e.g.,  with \ac{grams} \citep{2020APh...114..107A} or e-ASTROGAM \citep{2018JHEAp..19....1D}. If these observations will confirm the
two-component composition of the Crab Nebula spectrum, then one can attempt to define the magnetic field strength in the
``Maxwellian region''.  Although from Fig.~\ref{fig:maxwell} it may look as the \ac{lhaaso} sensitivity is not good
enough for such measurements, we remind that the shown sensitivity corresponds to one-year exposure. If the instrument
will operate long enough, e.g. ten years, its performance may appear to be sufficient for obtaining a meaningful
constraint on the magnetic field strength in this hypothetical ``Maxwellian region.''

\subsection{On the magnetization of the pulsar wind}

Presently, the \ac{mhd} treatment provides the most fruitful approach for studying the properties of \acp{pwn} \citep{Kennel1984b,2002MNRAS.336L..53B,2004MNRAS.349..779K,2005MNRAS.358..705B,2008A&A...485..337V,Camus2009,2014AN....335..234B,2014MNRAS.438..278P,2019MNRAS.484.4760B}.

The \ac{mhd} framework provides important insights into non-thermal physical processes in \acp{pwn}. In particular,
this concerns the dynamics of the magnetic field, particle transport and their radiation. Although the simplest \ac{1d} analytic
models helped to advance the studies  of \acp{pwn}, the most realistic results are achieved with
numerical \ac{3d} \ac{mhd} simulations \citep[see, e.g.,][]{2014MNRAS.438..278P,2019MNRAS.484.4760B}.

The \ac{mhd} models have limitations among which the phenomenological treatment of particle acceleration is essential. Although it is proved that relativistic outflows on different astrophysical scales are characterized by particle acceleration and radiation, \acp{pwn} demonstrate unprecedentedly high efficiency of non-thermal processes. Despite the systematic study of \acp{pwn}, it is still not fully understood  what makes of \acp{pwn} so efficient high-energy sources.

One of the key parameters in \ac{mhd} models applied to  \acp{pwn} is the magnetization of the pulsar wind, \(\sigma\), which determines the fraction of the pulsar spin-down losses that carried by the Poynting flux.  This parameter determines the magnetic field at the pulsar wind \ac{ts}. The downstream magnetic field at the \ac{ts} is
\be\label{eq:bfield}
B\simeq h(\sigma)\sqrt{\frac{L\sub{sd}}{ cR\sub{ts}^2}}\simeq 400 h(\sigma)\left(\frac{R\sub{ts}}{0.1\rm\ pc}\right)^{-1}\rm\,\upmu G\,,
\ee
where the function \(h\) accounts for the the Rankie-Hugoniot conditions at the \ac{ts} and the strength of the magnetic field in the unshocked pulsar wind:
 \(h\simeq1\) for \(1>\sigma\geq0.1\), and \(h(\sigma)\simeq3\sigma^{\nicefrac12}\) for \(\sigma<0.1\). In the case of the Crab Nebula, the radius of the termination shock is constrained robustly, \(R\sub{ts}\lesssim0.1\rm\,pc\), with the observations in the X-ray band \citep{Weisskopf2000}, the magnetic field at the \ac{ts} should exceed \(100\rm\,\upmu G\), unless the wind magnetization is very small, \(\sigma\leq10^{-2}\), or magnetic field dissipates at the \ac{ts} \citep{2003MNRAS.345..153L,2011ApJ...741...39S}.

As shown in Fig.~\ref{fig:syn_different_bfield}, the strength of the magnetic field in the region responsible for acceleration of the ``wind electrons'' should not exceed \(125\rm\,\upmu G\). Eq.~\eqref{eq:bfield}  shows that such a modest magnetic field requires very weak magnetization of the pulsar wind, \(\sigma \leq10^{-2}\). Although \ac{1d} \ac{mhd} models of the Crab Nebula, does require such a weakly magnetized pulsar wind,  currently, it is considered as an artefact of the ideal \ac{1d} approximation. Indeed, the rigid flow structure implemented in ideal \ac{1d} models results in a rapid increase of the magnetic field in the shocked pulsar wind. The initially weak magnetic field in the flow approaches the equipartition strength on the scale of several termination shock distances. For the magnetization of \(\sigma=3\times10^{-3}\), simple \ac{1d} models can reproduce the radiation spectrum \citep{Kennel1984b}, expansion rate, and, to some extent, also the X-ray morphology  band \citep{2002AstL...28..373B}.

\ac{3d} \ac{mhd} models agree better with the features of the Crab Nebula if one adopts a higher wind magnetization. As revealed by numerical simulations, the initially strong magnetic field can dissipate significantly in the shocked pulsar wind, allowing \ac{mhd} solutions with highly magnetized pulsar winds, \(\sigma\sim0.5\). However, such a strong magnetization implies  strong magnetic field at the \ac{ts}, \(B\simeq400\rm\,\upmu G\). The estimate based on  Eq.~\eqref{eq:bfield} is consistent with \ac{3d} numerical simulations by \citet{2014MNRAS.438..278P}. If \ac{uhe} electrons were accelerated in a region with such a strong magnetic field, their synchrotron emission would violate the level of the MeV flux. In  Fig.~11  (right panel) of \citet{2014MNRAS.438..278P} one can see a small region close in the equator plane characterized by a relatively weak magnetic field. Actually,  the equatorial region has been suggested as the most plausible site for the acceleration of TeV electrons in the Crab Nebula  \citep{2011ApJ...741...39S,2015MNRAS.449.3149O}, but a highly anisotropic pulsar wind might be required to supply enough energy to this relatively compact region.

\section{Conclusion}
The energy-dependent morphology seen in the center part of the Crab Nebula suggests that the TeV electrons  originate at the pulsar wind \ac{ts}. The \ac{ic} emission of these \ac{vhe} electrons smoothly extends to the \ac{uhe} regime as shown by the recent \ac{tibet} measurements.   Since \ac{uhe} electrons predominately interact with \ac{cmbr} photons, their spectrum and the total energetics is derived model-independently.  Because of the short cooling time, these electrons are confined in the proximity of the accelerator. The joint analysis of the fluxes of synchrotron and \ac{ic} components reveals a weak magnetic field, $\leq120 \rm\, \upmu G$, in the acceleration site responsible for acceleration of multi-hundred TeV electrons in the Crab Nebula. To obtain such a weak magnetic field at the pulsar wind \ac{ts}, one needs either to assume a small magnetization of the pulsar wind, \(\sigma\leq10^{-2}\) or a highly anisotropic pulsar wind and efficient magnetic field reconnection operating at the \ac{ts}/bulk of the nebula. Either of these possibilities needs to be tested against realistic \ac{3d} \ac{mhd} simulations of the Crab Nebula.

The obtained limitation on the magnetic field might be not valid in the region responsible for acceleration of PeV electrons in the Crab Nebula. The future observations above \(1\rm\,MeV\) and \(300\rm\,TeV\) will reveal the physical conditions in that region. In particular, these observations have a potential  (1) to verify if the broad band spectrum requires a presence of an additional electron component with a narrow energy distribution, and (2) to constrain the magnetic field strength in the region responsible for acceleration of the PeV electrons.

If this future study favors a strong magnetic field in the region of acceleration of PeV electrons, which are responsible
for the \(\sim100\rm\,MeV\) steady emission in the Crab Nebula, then the site(s) of acceleration of this component could
be responsible also for the enhanced GeV emission observed during the Crab flares.  Thus, the future observations of \ac{uhe} gamma rays up to \(1\rm\, PeV\) may shed light on the origin of this
component, and, perhaps also, its links to the Crab flares.

\section*{Acknowledgements}
The authors thank V.Bosch-Ramon and anonymous referee for their useful comments. DK is supported by JSPS KAKENHI Grant Numbers JP18H03722, JP24105007, and JP16H02170. M.A. is supported by the RIKEN Junior Research Associate Program. 








\appendix

\section{Markov Chain Monte Carlo modelling of the hard X-ray and multi-TeV emission from the Crab Nebula}
\label{sec:MCMC}
To study the constraints imposed by the \ac{uhe} gamma-ray spectrum measured from the Crab Nebula, we fitted the \ac{integral} and gamma-ray (above \(10\rm \,TeV\)) spectra with a synchrotron~--~\ac{ic} model. The energy distribution of emitting particles is assumed to be a power-law with exponential cutoff:
\be\label{eq:el_mcmc}
\dif[E]{N} = A \left(\frac{E}{1\rm\,TeV}\right)^{-\alpha} \exp{\left[-\left(\frac{E}{E_{\rm cut}}\right)^\beta\right]}\,,
\ee
where \(E\) is electron energy. The particle spectrum is determined by  four 
parameters: \(A\) is normalization;
\(E_{\rm cut}\) is cutoff energy; $\alpha$ is the spectral index of a power-law distribution; and \(\beta\) is the
cutoff index.
Leptons produce X-ray emission through synchrotron radiation in a magnetic field, which is assumed to have
a random orientation but uniform strength \(B\).  The gamma-ray emission is generated through IC scattering on
\ac{cmbr}, \ac{fir} (a graybody distribution with temperature $T\sub{fir}=$ 70 K and energy density $U\sub{fir}=$ 0.5
$\rm{eV~cm^{-3}}$), \ac{nir} ($kT\sub{nir}=$ 5000 K, $U\sub{nir}=$ 1 $\rm{eV~cm^{-3}}$), and synchrotron photons
(assuming a homogeneously generated target). We note that the \ac{cmbr} photons provide the dominant
contribution, so our specific assumptions about the photon targets have only a minor influence on the result.

The computation of the \ac{sed} models and subsequent fit to the multiwavelength \acp{sed} are performed using the
{\tt naima} Python package \citep{naima}. Specifically, synchrotron emission is computed based on the formalism in
\citet{2010PhRvD..82d3002A}, and \ac{ic} emission on the formalism in \citet{1981Ap&SS..79..321A} and
\citet{2014ApJ...783..100K}. {\tt naima} allows one to obtain the best-fit values and posterior probability
distributions of the model parameters given the \ac{sed} points from the $\chi^2$, calculated assuming that the \ac{sed}
point uncertainties are Gaussian and uncorrelated.  The model parameters are scanned using the \ac{mcmc} method, as
implemented in the \texttt{emcee} package \citep{foreman2013}. For all model parameters we assume a flat prior
probability distribution, within physical constraints on the parameters values (e.g., particle densities are
positive). 
We scan the normalization $A$ and the cutoff energy $E_{\rm cut}$ in logarithmic space, so that the fit parameter is actually
$\log_{10}(A/\mathrm{eV}^{-1})$ and $\log_{10}(E_{\rm cut}/1\,\mathrm{TeV})$, respectively.

\begin{figure*}
  \includegraphics[width=2\columnwidth]{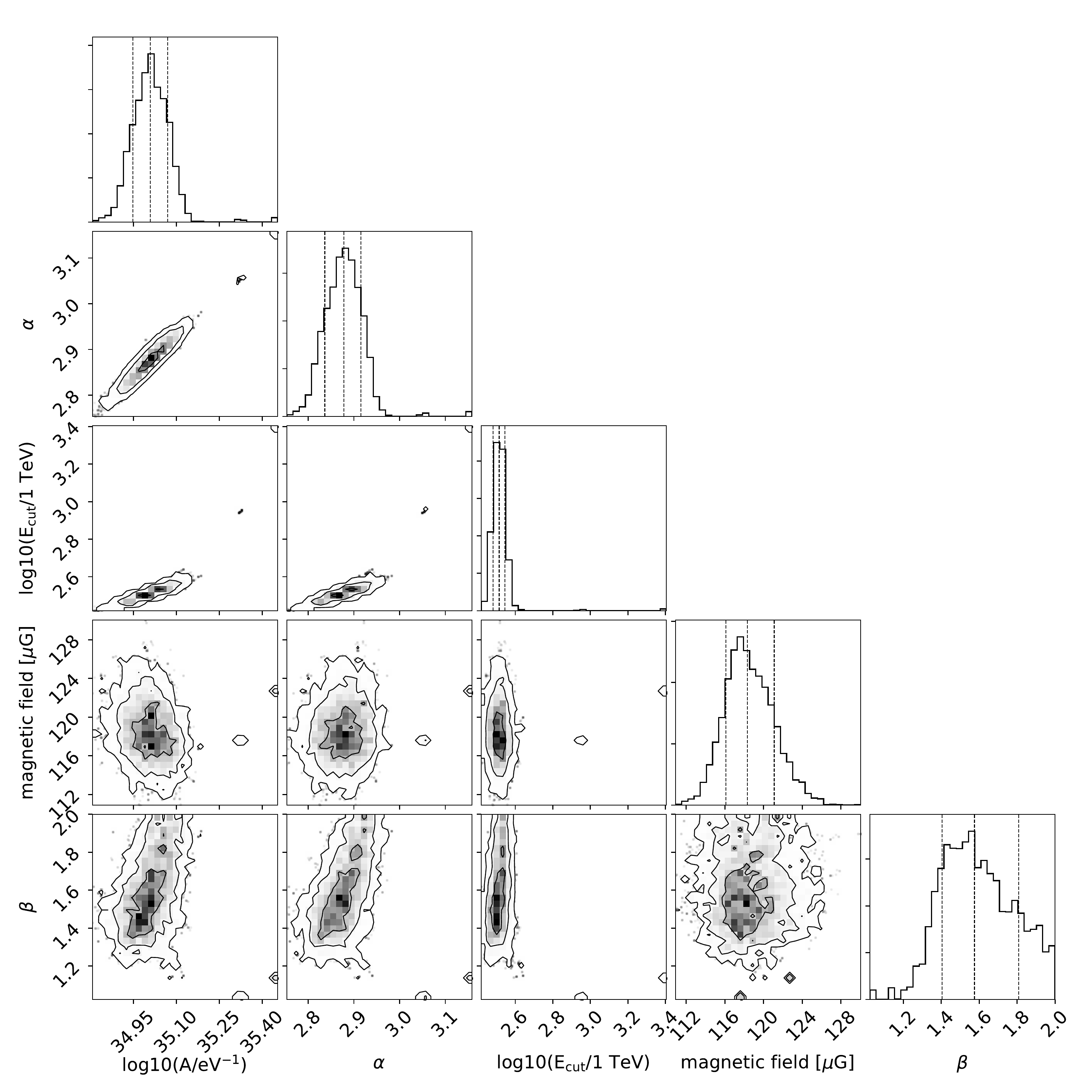}
  \caption{One- and two-dimensional projections of the posterior probability density distributions of the parameters for the radiative model for the hard X-ray and gamma-ray (above \(10\rm\,TeV\)) spectra of the Crab Nebula. The parameters of the electron spectrum are defined by Eq.~\ref{eq:el_mcmc}.
  The lines overlaid to the one-dimensional projections are the 16th, 50th and 84th percentiles of the distributions. The contours overlaid to the two-dimensional projections correspond to $1\sigma$, $1.5\sigma$, and $2\sigma$ probability decrease with respect to the maximum. \label{fig:mcmc}}
\end{figure*}

The fit results are shown in Fig.~\ref{fig:mcmc} and summarized in Table~\ref{tab:mcmc}. Figure~\ref{fig:mcmc} contains
the probability density distributions of the model parameters, and Table~\ref{tab:mcmc} gives median and upper and lower
uncertainties on the parameter values. The fitting shows that the hard X-ray and multi-TeV data constrain the magnetic
field strength with very high accuracy. As there could be regions with stronger magnetic field in the nebula, lower
energy electrons may provide important contribution to the hard X-ray band without producing any sensible \ac{ic}
emission. Thus, one should rather  take the \ac{integral} measurements as upper limits for the X-ray emission from the zone,
where \ac{uhe} electrons are accelerated, i.e., \(B<120\,\rm\upmu G\) in that region.

\begin{table}
	\centering
	\caption{Results of fitting of the hard X-ray and multi-TeV spectra of the Crab Nebula with {\tt naima}.}
	\label{tab:mcmc}
	\begin{tabular}{lc} 
          \hline
          parameter & value \\
          \hline
          magnetic field & \(B=118_{-2}^{+3}\rm\,\upmu G\)\\
          normalization & \(A=1.0_{-0.1}^{+0.2}\times10^{35}\)\rm\,eV$^{-1}$\\
          power-law index & \(\alpha=2.88\pm0.04\)\\
          cutoff energy& \(E_{\rm cut}= 330\pm20\rm\,TeV\)\\
          cutoff index& \(\beta= 1.6\pm0.2\)\\
          \hline
	\end{tabular}
\end{table}

\bsp	
\label{lastpage}
\end{document}

%% file: mycommands.tex
\usepackage{pgffor}
\makeatletter
\def\dif{\@ifnextchar[{\@with}{\@without}}
\def\@with[#1]#2{
  \ensuremath{\frac{\foreach \x in {#2}{\mathrm{d}\x\,}}{\foreach \x in {#1}{\mathrm{d}\x\,}}}
}
\def\@without#1{
  \ensuremath{%
    \ifx\hfuzz#1\hfuzz
    \mathrm{d}
    \else
    \foreach \x in {#1}{\mathrm{d}\x\,}
    \fi
    }
}
\makeatother

\newif\iflong
\longfalse

\newcommand{\ergs}{\ensuremath{\mathrm{erg\,s^{-1}}}}

\newcommand{\flux}{\ensuremath{\mathrm{erg\,s^{-1}\,cm^{-2}}}}

\newcommand{\be}{\begin{equation}}
\newcommand{\ee}{\end{equation}}
\newcommand{\ba}{\begin{eqnarray}}
\newcommand{\ea}{\end{eqnarray}}

\newcommand{\ve}{\ensuremath{\varepsilon}}

\newcommand{\sub}[1]{_\textsc{#1}}

\newcommand{\syn}{_\textsc{syn}}

\newcommand{\dias}{{Dublin Institute for Advanced Studies, School of Cosmic Physics, 31 Fitzwilliam Place, Dublin 2, Ireland}}
\newcommand{\mephi}{{National Research Nuclear University (MEPHI), Kashirskoje shosse, 31, Moscow 115409, Russia}}
\newcommand{\mpik}{{Max-Planck-Institut f\"ur Kernphysik, Saupfercheckweg 1, 69117 Heidelberg, Germany}}
\newcommand{\rikkyo}{{Department of Physics, Rikkyo University, Nishi-Ikebukuro 3-34-1, Toshima-ku, Tokyo 171-8501, Japan}}

%% file: myacro_arxiv.tex
\usepackage[single=true]{acro}
\DeclareAcronym{agn}{
  short = AGN,
  long  = active galactic nucleus,
  short-plural = s,
  long-plural-form = active galactic nuclei,
  class = astro,
  first-style = default
}
\DeclareAcronym{smbh}{
  short = SMBH ,
  long  = supermassive black hole ,
  class = astro ,
  first-style = default
}
\DeclareAcronym{pwn}{
  short = PWN ,
  long  = pulsar wind nebula ,
  short-plural = e ,
  long-plural = e ,
  class = astro ,
  first-style = default
}
\DeclareAcronym{cr}{
  short = CR ,
  long  = cosmic ray ,
  class = astro ,
  first-style = default
}
\DeclareAcronym{ns}{
  short = NS ,
  long  = neutron star ,
  class = astro ,
  first-style = default
}
\DeclareAcronym{ism}{
  short = ISM ,
  long  = interstellar medium ,
  class = astro ,
  first-style = default
}

\DeclareAcronym{cmbr}{
  short = CMBR ,
  long  = Cosmic Microwave Background Radiation ,
  class = astro,
  first-style = default
}
\DeclareAcronym{ebl}{
  short = EBL ,
  long  = extragalactic background light ,
  class = astro,
  first-style = default
}
\DeclareAcronym{fir}{
  short = FIR ,
  long  = far-infrared ,
  class = astro,
  first-style = default
}
\DeclareAcronym{nir}{
  short = NIR ,
  long  = near-infrared ,
  class = astro,
  first-style = default
}
\DeclareAcronym{dsa}{
  short = DSA ,
  long  = diffusive shock acceleration ,
  class = astro ,
  first-style = default
}
\DeclareAcronym{sto}{
  short = StA ,
  long  = stochastic acceleration ,
  class = astro ,
  first-style = default
}
\DeclareAcronym{ic}{
  short = IC ,
  long  = inverse Compton ,
  class = astro ,
  first-style = default
}
\DeclareAcronym{ssa}{
  short = SSA ,
  long  = synchrotron-self-abosrption ,
  class = astro ,
  first-style = default
}
\DeclareAcronym{ssc}{
  short = SSC ,
  long  = synchrotron-self-Compton ,
  class = astro ,
  first-style = default
}
\DeclareAcronym{kn}{
  short = KN ,
  long  = Klein~--~Nishina ,
  class = astro ,
  first-style = default
}
\DeclareAcronym{ec}{
  short = EC ,
  long  = elastic Coulomb ,
  class = astro ,
  first-style = default
}
\DeclareAcronym{pp}{
  short = \ensuremath{pp} ,
  long  = hadronuclear ,
  class = astro ,
  first-style = default
}
\DeclareAcronym{pg}{
  short = \ensuremath{p\gamma} ,
  long  = photomeson ,
  class = astro ,
  first-style = default
}

\DeclareAcronym{he}{
  short = HE ,
  short-plural-form = HE ,
  long  = high-energy ,
  long-plural-form = high energies ,
  class = hydro ,
  first-style = default
}
\DeclareAcronym{vhe}{
  short = VHE ,
  short-plural-form = VHE ,
  long  = very high-energy ,
  class = astro ,
  first-style = default 
}
\DeclareAcronym{uhe}{
  short = UHE ,
  short-plural-form = UHE ,
  long  = ultra high-energy ,
  class = astro ,
  first-style = default 
}
\DeclareAcronym{sed}{
  short = SED ,
  long  = Spectral Energy Distribution ,
  single  = spectral energy distribution ,
  class = astro ,
  first-style = default
}
\DeclareAcronym{mec}{
  short = MEC ,
  long  = monochromatic emission coefficient ,
  class = astro ,
  first-style = default
}

\DeclareAcronym{hd}{
  short = HD ,
  long  = hydrodynamic ,
  class = hydro ,
  first-style = default 
}

\DeclareAcronym{mhd}{
  short = MHD ,
  long  = magnetohydrodynamic ,
  class = hydro ,
  first-style = default 
}
\DeclareAcronym{rmhd}{
  short = RMHD ,
  long  = relativistic magnetohydrodynamic ,
  class = hydro ,
  first-style = default
}
\DeclareAcronym{1d}{
  short = 1D ,
  long  = one-dimensional ,
  class = hydro ,
  first-style = short
}
\DeclareAcronym{2d}{
  short = 2D ,
  long  = two-dimensional ,
  class = hydro ,
  first-style = short
}
\DeclareAcronym{3d}{
  short = 3D ,
  long  = three-dimensional ,
  class = hydro ,
  first-style = short
}
\DeclareAcronym{cd}{
  short = CD ,
  long  = contact discontinuity ,
  class = hydro ,
  first-style = default
}
\DeclareAcronym{ts}{
  short = TS ,
  long  = termination shock ,
  class = hydro ,
  first-style = default
}
\DeclareAcronym{eos}{
  short = EOS ,
  long  =  equation of state ,
  long-plural-form = equations of state ,
  class = hydro ,
  first-style = default
}

\DeclareAcronym{pic}{
  short = PIC ,
  long = particle-in-cell ,
  first-style = default
}
\DeclareAcronym{mcmc}{
  short = MCMC ,
  long = Markov~Chain Monte~Carlo ,
  first-style = default
}

\DeclareAcronym{mq}{
  short = \(\upmu\)Q ,
  long = microquasar ,
  first-style = default
}

\DeclareAcronym{alma}{
  short = ALMA ,
  long  = Atacama Large Millimeter/submillimeter Array ,
  class = instruments ,
  first-style = default
}
\DeclareAcronym{nustar}{
  short = {\it NuSTAR} ,
  long  = Nuclear Spectroscopic Telescope Array ,
  class = instruments ,
  first-style = default
}
\DeclareAcronym{integral}{
  short = {\it INTEGRAL} ,
  long  = {INTErnational Gamma-Ray Astrophysics Laboratory} ,
  single = {International Gamma-Ray Astrophysics Laboratory} ,
  class = instruments ,
  first-style = default
}
\DeclareAcronym{comptel}{
  short = {\it COMPTEL} ,
  long  = Imaging Compton Telescope ,
  class = instruments ,
  first-style = default
}
\DeclareAcronym{grams}{
  short = {\it GRAMS} ,
  long  = Gamma-Ray and AntiMatter Survey ,
  class = instruments ,
  first-style = default
}

\DeclareAcronym{fermi}{
  short = {\it Fermi}/LAT ,
  long  = {\it Fermi} Large Area Telescope ,
  class = instruments ,
  first-style = default
}
\DeclareAcronym{hegra}{
  short = HEGRA ,
  long  = High Energy Gamma Ray Astronomy ,
  class = instruments ,
  first-style = default
}
\DeclareAcronym{tibet}{
  short = {\it Tibet AS\(\gamma\)} ,
  long  = Tibet Air Show Gamma Experiment,
  class = instruments ,
  first-style = default
}
\DeclareAcronym{hawc}{
  short = {\it HAWC} ,
  long  = High Altitude Water Cherenkov Gamma-Ray Observatory ,
  class = instruments ,
  first-style = default
}

\DeclareAcronym{lhaaso}{
  short = {\it LHAASO} ,
  long  = Large High Altitude Air Shower Observatory ,
  class = instruments ,
  first-style = default
}